\documentclass[12pt,twoside]{article}

\usepackage{graphicx}
\usepackage{epsfig}
\usepackage{url}
\usepackage{psfrag}
\usepackage{times}
\usepackage{amsfonts,amssymb,amsbsy}
\usepackage{amsmath}
\usepackage{latexsym}
\usepackage{amsfonts,amsbsy}
\usepackage{latexsym}

\begin{document}

\def\e{\begin{equation}} 
\def\f{\end{equation}} 
\def\ea{\begin{eqnarray}} 
\def\fa{\end{eqnarray}} 

\def\##1{{\mbox{\textbf{#1}}}}
\def\%#1{{\mbox{\boldmath $#1$}}}
\def\=#1{{\overline{\overline{\mathsf #1}}}}

\def\*{^{\displaystyle*}}
\def\xx{\displaystyle{{}^\times}\llap{${}_\times$}}
\def\.{\cdot}
\def\x{\times}
\def\oo{\infty}

\def\D{\nabla}
\def\d{\partial}

\def\ra{\rightarrow}
\def\lra{\leftrightarrow}
\def\Ra{\Rightarrow}
\def\le{\left(}
\def\ri{\right)}
\def\l#1{\label{eq:#1}}
\def\r#1{(\ref{eq:#1})}
\def\am{\left(\begin{array}{c}}
\def\amm{\left(\begin{array}{cc}}
\def\ammm{\left(\begin{array}{ccc}}
\def\ammmm{\left(\begin{array}{cccc}}
\def\a{\end{array}\right)}

\def\I{\int\limits}
\def\OI{\oint\limits}

\def\A{\alpha}
\def\B{\beta}
\def\de{\delta}
\def\De{\Delta}
\def\E{\epsilon}
\def\g{\gamma}
\def\G{\Gamma}
\def\h{\eta}
\def\K{\kappa}
\def\la{\lambda}
\def\La{\Lambda}
\def\M{\mu}
\def\o{\omega}
\def\Om{\Omega}
\def\R{\rho}
\def\s{\sigma}
\def\t{\tau}
\def\z{\zeta}
\def\X{\chi}
\def\TH{\theta}
\def\Th{\Theta}
\def\VF{\varphi}
\def\VR{\varrho}
\def\VT{\vartheta}
\def\ve{\%\varepsilon}


\title{Electromagnetic Wave Reflection from\\ Surface with General Boundary Conditions}
\author{I.V. Lindell and A. Sihvola} 
\date{School of Electrical Engineering,\\ Aalto University, Espoo, Finland\\ 
{\tt ismo.lindell@aalto.fi}\\\vspace{-1pt}{\tt ari.sihvola@aalto.fi}}
\pagestyle{myheadings}

\pagestyle{myheadings}

\textwidth=16cm
\hoffset=-1.5cm
\voffset=-2cm
\textheight=24cm
\parindent=0pt
\parskip=\medskipamount

\maketitle

\begin{abstract}
The most general linear and local set of boundary conditions, involving relations between the normal components of the D and B vectors and tangential components of the E and H vectors at each point of the boundary, are considered in this paper. Reflection of a plane wave from a boundary defined by general conditions in an isotropic half space is analyzed and an analytic expression for the reflection dyadic is derived. It is shown that any plane wave can be decomposed in two components which do not interact in reflection. Properties of plane waves matched to the general boundary are given. Certain special cases of boundary conditions, arising naturally from the general theory and labeled as E-boundary, H-boundary and EH-boundary conditions, are introduced as interesting novelties and some of their properties are studied. Previously known special cases are considered in verifying the theory. A possible realization of the general boundary in terms of an interface of a general bi-anisotropic medium is discussed in an Appendix. 
\end{abstract}

\section{Introduction}

To make solutions for electromagnetic problems unique, boundary conditions at the surface bounding the region of interest are needed. If the surface defines an interface between two regions, a set of interface conditions is required to match the fields of both sides of the interface. In the literature, boundary conditions are often applied to approximate material interfaces, e.g., when solving scattering problems \cite{Senior,Hoppe}. Replacing the interface by a clever set of approximate boundary conditions reduces the region of computation and, hopefully, does not import too much error. Here we consider boundary conditions which are exact by definition. The problem is then to realize the medium interface so that the boundary conditions are satisfied as well as possible. A structure defined by a surface with given boundary conditions is often designed to satisfy a certain purpose, for example, to obtain an antenna with given radiation properties. Realization of the antenna calls for realization of the boundary by certain material or metamaterial. As a concrete example, realization of the perfect magnetic conductor (PMC) boundary has been a challenge to antenna engineering \cite{Sievenpiper}. 

The present paper considers properties of boundaries defined by linear boundary conditions of very general form. The conditions are assumed to be local, i.e., the field relations do not depend on spatial variation of the fields. One must notice here that, in contrast, the most effective approximate boundary conditions are nonlocal, defined by boundary parameters containing differential operators \cite{Senior,Hoppe}.

\subsection{General Boundary Conditions}

In \cite{PIERL2016} we have shown that the most general linear and local boundary conditions can be expressed in the form
\e \A_1 c\#n\.\#B+ \frac{\B_1}{\E_o}\#n\.\#D+ \#a_{1t}\.\#E+ \h_o\#b_{1t}\.\#H=0, \l{ximp1}\f
\e \A_2 c\#n\.\#B+ \frac{\B_2}{\E_o}\#n\.\#D+ \#a_{2t}\.\#E+ \h_o\#b_{2t}\.\#H=0, \l{ximp2}\f
where $\#n$ denotes the unit vector normal to the boundary surface and vectors with subscript $()_t$ are tangential to the boundary:
\e \#n\.\#a_{1t}=\#n\.\#a_{2t}=\#n\.\#b_{1t}=\#n\.\#b_{2t}=0. \f 
The coefficients 
\e c = 1/\sqrt{\M_o\E_o},\ \ \ \ \h_o=\sqrt{\M_o/\E_o}, \f
and $1/\E_o$ have been added to the definitions of \cite{PIERL2016} to make the scalars $\A_1,\A_2,\B_1,\B_2$ and the vectors  $\#a_{1t},\#a_{2t},\#b_{1t},\#b_{2t}$  dimensionless. Altogether, the parameters involve $4+4\x2=12$ parameters. However, since the two conditions \r{ximp1}, \r{ximp2} are homogeneous, the actual number of parameters defining the boundary is 10.

\subsection{Special Cases}

The general form of boundary conditions \r{ximp1} and \r{ximp2} can be seen as the final stage of a process which was started by us from the simple DB boundary \cite{DB} through various generalizating steps (SHDB and GSHDB) \cite{SHDB,GSHDB}, with increasing number of parameters. The different stages can be recovered by considering various special cases of the general conditions, for specific choices of the four vectors and four scalars in \r{ximp1} and \r{ximp2}. 

The generalized soft-and-hard/DB (GSHDB) boundary \cite{GSHDB} is obtained by substituting $\B_1=0$, $\A_2=0$, $\#b_{1t}=0$ and $\#a_{2t}=0$ in \r{ximp1} and \r{ximp2}. The GSHDB boundary conditions involve two scalars and two 2D vectors,
\e \A_1 c\#n\.\#B+ \#a_{1t}\.\#E=0, \l{GSHDB1}\f
\e \B_2 c\#n\.\#D+ \#b_{2t}\.\#H=0 \l{GSHDB2}\f
which requires four free parameters for the definition.
 
As a special case of the GSHDB boundary, the soft-and-hard/DB (SHDB) boundary \cite{SHDB} is defined by
\e \A_1 c\#n\.\#B+ \#a_{1t}\.\#E=0, \l{SHDB1}\f
\e \A_1 c\#n\.\#D- \#a_{1t}\.\#H=0, \l{SHDB2}\f
whose special cases are the soft-and-hard (SH) boundary \cite{SH},
\e \#a_{1t}\.\#E=0,\ \ \ \ \ \#a_{1t}\.\#H=0, \l{SH}\f
which has the generalization (GSH) \cite{GSH},
\e \#a_{1t}\.\#E=0,\ \ \ \ \ \#b_{2t}\.\#H=0, \l{GSH}\f
and the DB boundary \cite{DB}, defined by
\e \#n\.\#D=0,\ \ \ \ \ \#n\.\#B=0. \l{DB}\f
Properties of waves reflecting from boundaries obeying these conditions have been recently studied. For example, it has been shown that in all of the above cases the incident field can be decomposed in two components, one of which is reflected as from the perfect electric conductor (PEC) boundary defined by $\#n\x\#E=0$, and, the other one, as from the perfect magnetic conductor (PMC) boundary defined by $\#n\x\#H=0$.

The impedance-boundary conditions can be obtained as another special case of \r{ximp1}, \r{ximp2}, by assuming $\A_1=\A_2=\B_1=\B_2=0$. This leaves four 2D vectors for the definition,
\e \#a_{1t}\.\#E+ \h_o\#b_{1t}\.\#H=0, \l{imp1}\f
\e  \#a_{2t}\.\#E+ \h_o\#b_{2t}\.\#H=0. \l{imp2}\f
\r{imp1} and \r{imp2} can be rewritten in the form of a 2D vector condition \cite{Methods},
\e \#E_t=\=Z_s\.(\#n\x\#H), \f
where $\=Z_s$ is the surface-impedance dyadic satisfying $\#n\.\=Z_s=\=Z_s\.\#n=0$.

The purpose of \cite{PIERL2016}, where the general conditions \r{ximp1}, \r{ximp2} were introduced, was to show that if such a boundary is required to have the property that one of its two eigenwaves is reflected as from the PEC boundary, and the other one as from the PMC boundary, it must be equal the generalized soft-and-hard/DB (GSHDB) boundary. 

It is the purpose of the present paper to study other properties of the general boundary, including decomposition of the incident wave in two components which retain their properties after reflection, and definition of waves matched to the general boundary. The general form allows us to study some obvious but unconventional special cases of the general boundary. Finally, in an Appendix, a possible realization of the boundary defined by \r{ximp1}, \r{ximp2}, as an interface of a bi-anisotropic medium is sketched.

\section{Reflection from Boundary}

 In the following, we assume for simplicity that the boundary surface defined by $\#n\.\#r=0$ is planar as defined by a constant unit vector $\#n$. Also, a simple isotropic medium is assumed in the half space $\#n\.\#r>0$, as defined by the parameters $\E_o,\M_o$. In this case, the boundary conditions \r{ximp1} and \r{ximp2} can be expressed in the more compact form
\ea  \#a_1\.\#E + \h_o \#b_1\.\#H &=& 0, \l{aEbH1}\\
       \#a_2\.\#E + \h_o \#b_2\. \#H &=&0, \l{aEbH2}\fa
with
\e \#a_j = \B_j \#n + \#a_{jt},\ \ \ \ \ \#b_j = \A_j\#n + \#b_{jt},\ \ \ \ \ j=1,2. \f
\r{aEbH1} and \r{aEbH2} actually generalize the impedance-boundary conditions \r{imp1} and \r{imp2} by adding normal components to the vectors $\#a_{jt}$ and $\#b_{jt}$. However, one must keep in mind that \r{aEbH1} and \r{aEbH2} correspond to the conditions \r{ximp1} and \r{ximp2} only when the medium above the boundary is isotropic.

\subsection{Plane-Wave Properties}

Let us consider a time-harmonic plane wave incident to the boundary and reflecting from it. The electric fields are defined by 
\e \#E^i(\#r)= \#E^i\exp(-j\#k^i\.\#r),\ \ \ \ \#E^r(\#r)= \#E^r\exp(-j\#k^r\.\#r), \f
with wave vectors satisfying
\e \#k^i = \#k_t -k_n\#n,\ \ \ \ \#k^r=\#k_t + k_n\#n, \f
\e \#k^i\.\#k^i=\#k^r\.\#k^r=k_o^2 = \o^2\M_o\E_o.\f
Applying the relations 
\e \#k^i\x\#E^i = k_o\h_o\#H^i,\ \ \ \#k^i\x\h_o\#H^i = -k_o\#E^i,  \l{Maxi}\f
\e \#k^r\x\#E^r = k_o\h_o\#H^r,\ \ \ \#k^r\x\h_o\#H^r = -k_o\#E^r, \l{Maxr}\f
following from the Maxwell equations, the boundary conditions \r{aEbH1} and \r{aEbH2} for the fields of the sum of incident and reflected waves take the form
\ea k_o\#a_1\.(\#E^i+\#E^r) + \#b_1\.(\#k^i\x\#E^i+ \#k^r\x\#E^r) &=& 0, \\
     k_o\#a_2\.(\#E^i+\#E^r) + \#b_2\.(\#k^i\x\#E^i+ \#k^r\x\#E^r) &=& 0. \fa
These can be rewritten as
\ea \#c_1^i\.\#E^i + \#c_1^r\.\#E^r &=& 0, \l{c1E}\\
     \#c_2^i\.\#E^i + \#c_2^r\.\#E^r &=& 0, \l{c2E}\fa
by introducing the four vectors
\e \#c_1^i = \#k^i\x\#b_1 -k_o\#a_1,\ \ \ \ \#c_1^r = \#k^r\x\#b_1 - k_o\#a_1, \l{c1}\f
\e \#c_2^i = \#k^i\x\#b_2 -k_o\#a_2,\ \ \ \ \#c_2^r = \#k^r\x\#b_2 - k_o\#a_2. \l{c2}\f
Defining
\e J^i = \#k^i\.\#c_1^i\x\#c_2^i, \ \ \ \ J^r=\#k^r\.\#c_1^r\x\#c_2^r, \l{JiJr}\f
and assuming $J^i\not=0, J^r\not=0$, we can form two bases of vectors  $\{\#k^i,\ \#c_1^i,\ \#c_2^i\}$ and $\{\#k^r,\ \#c_1^r,\ \#c_2^r\}$, with their reciprocal vector bases respectively defined by \cite{Gibbs}
\e  \{\frac{1}{J^i}\#c_1^i\x\#c_2^i,\ \ \ \frac{-1}{J^i}\#k^i\x\#c_2^i,\ \ \ \frac{1}{J^i}\#k^i\x\#c_1^i\}\f
\e \{\frac{1}{J^r}\#c_1^r\x\#c_2^r,\ \ \ \frac{-1}{J^r}\#k^r\x\#c_2^r,\ \ \ \frac{1}{J^r}\#k^r\x\#c_1^r\}.\f
The unit dyadic can be expressed in terms of either set of basis vectors as

\ea \=I&=&\frac{1}{J^i}((\#c_1^i\x\#c_2^i)\#k^i -(\#k^i\x\#c_2^i)\#c_1^i + (\#k^i\x\#c_1^i)\#c_2^i) \\
&=& \frac{1}{J^r}((\#c_1^r\x\#c_2^r)\#k^r -(\#k^r\x\#c_2^r)\#c_1^r + (\#k^r\x\#c_1^r)\#c_2^r). \fa

Taking  $\#k^i\.\#E^i=\#k^r\.\#E^r=0$ into account, the incident and reflected electric fields can be expanded in terms of the respective basis vectors as
\ea \#E^i &=& \frac{1}{J^i}(-(\#k^i\x\#c_2^i)(\#c_1^i\.\#E^i) + (\#k^i\x\#c_1^i)(\#c_2^i\.\#E^i)), \l{EiEi}\\
\#E^r &=&  \frac{1}{J^r}(-(\#k^r\x\#c_2^r)(\#c_1^r\.\#E^r) + (\#k^r\x\#c_1^r)(\#c_2^r\.\#E^r)). \l{ErEr}\fa
Substituting \r{c1}, \r{c2} and \r{Maxi}, \r{Maxr} leads to the representations
\e \#E^i = \frac{k_o}{J^i}\#k^i\x(\#c_2^i(\#b_1\.\h_o\#H^i+ \#a_1\.\#E^i) - \#c_1^i(\#b_2\.\h_o\#H^i+\#a_2\.\#E^i)), \l{32}\f
\e\#E^r = \frac{k_o}{J^r}\#k^r\x(\#c_2^r(\#b_1\.\h_o\#H^r+ \#a_1\.\#E^r) - \#c_1^r(\#b_2\.\h_o\#H^r+\#a_2\.\#E^r)). \l{33}\f

\subsection{Reflection Dyadic}

Applying \r{c1E} and \r{c2E}, the expansions \r{EiEi} and \r{ErEr} yield two relations between the incident and reflected fields,
\ea \#E^i &=& \frac{1}{J^i}((\#k^i\x\#c_2^i)(\#c_1^r\.\#E^r) - (\#k^i\x\#c_1^i)(\#c_2^r\.\#E^r)), \l{EiEr}\\
\#E^r   &=& \frac{1}{J^r}((\#k^r\x\#c_2^r)(\#c_1^i\.\#E^i) - (\#k^r\x\#c_1^r)(\#c_2^i\.\#E^i)). \l{ErEi}\fa
One can easily verify that these are really inverse relations. Writing the latter in the form
\e \#E^r= \=R\.\#E^i, \l{ERE}\f
the following analytic expression for the reflection dyadic $\=R$ can be found,
\e \=R =  \frac{1}{\#k^r\.\#c_1^r\x\#c_2^r}\#k^r\x(\#c_2^r\#c_1^i - \#c_1^r\#c_2^i).  \l{R}\f

As a simple check, let us apply \r{R} to the PEC boundary defined by $\#b_1=\#b_2=0$ and $\#a_1=\#u_x,\ \#a_2=\#u_y$, where $\#u_x$ and $\#u_y$ are tangential orthogonal unit vectors with $\#n=\#u_x\x\#u_y=\#u_z$. From $\#c_1^r=\#c_1^i=-k_o\#u_x$ and  $\#c_2^r=\#c_2^i=-k_o\#u_y$, we can expand \r{R} as
\ea \=R &=& \frac{1}{k_n}\#k^r\x(\#u_y\#u_x - \#u_x\#u_y) \nonumber\\
&=& \frac{1}{k_n}\#k^r\x(\#n\x\=I_t) = -\=I_t + \frac{\#n\#k_t}{k_n}, \l{RPEC}\fa
with $\=I_t=\=I-\#n\#n$. This yields the known result for the field reflected from the PEC boundary, $\#E^r = \=R\.\#E^i = -\#E_t^i + \#n E_n^i. $ 

As another check, the soft-and-hard (SH) boundary can be defined by assuming $\#a_1= \#b_2=\#v_t$, $\#a_2=\#b_1=0$  with $\#n\.\#v_t=0$, whence  the boundary conditions \r{aEbH1} and \r{aEbH2} become
\e \#v_t\.\#E=\#v_t\.\#H=0.  \f
Substituting $\#c_2^r\#c_1^i-\#c_1^r\#c_2^i=-k_o(\#k^r\x\#v_t\#v_t - \#v_t\#k^i\x\#v_t)$, the reflection dyadic \r{R} is equivalent to the symmetric expression, 
$$ \=R = \frac{-1}{k_o^2(\#v_t\x\#k^r)^2}(k_o^2(\#k^r\x\#v_t)(\#k^i\x\#v_t)+ $$
\e + \#k^r\x(\#k^r\x\#v_t)\#k^i\x(\#k^i\x\#v_t)), \f
which coincides with a result previously derived in \cite{MAP1996}, eqn. (20).

\subsection{Wave decomposition}

From \r{EiEi} and \r{ErEr} one can see that both the incident field and the reflected field can be uniquely expressed in terms of two scalar potential-like quantities $\#c_1^i\.\#E^i, \#c_2^i\.\#E^i$ and $\#c_1^r\.\#E^r, \#c_2^r\.\#E^r$, respectively. This means that both waves can be decomposed in two parts as 
\e \#E^i = \#E^i_1 + \#E^i_2,\ \ \ \ \ \#E^r=\#E^r_1+ \#E^r_2, \f 
\e \#E^{i,r}_1 = \frac{1}{J^{i,r}}(\#k^{i,r}\x\#c_1^{i,r})\#c_2^{i,r}\.\#E^{i,r}, \f
\e \#E^{i,r}_2 = \frac{-1}{J^{i,r}}(\#k^{i,r}\x\#c_2^{i,r})\#c_1^{i,r}\.\#E^{i,r}. \f
They may be labeled as incident and reflected TE$_{c1}$ and TE$_{c2}$ waves because the electric fields satisfy
\e \#c_1^i\.\#E_1^i =\#c_2^i\.\#E_2^i =0, \f
\e \#c_1^r\.\#E_1^r =\#c_2^r\.\#E_2^r =0. \f
From
\e \#c_1^r\.\=R\.\#E^i = \frac{1}{\#k^r\.\#c_1^r\x\#c_2^r}\#c_1^r\.(\#k^r\x\#c_2^r)(\#c_1^i\.\#E^i) =-\#c_1^i\.\#E^i , \f
and, similarly, from $\#c_2^r\.\=R\.\#E^i=-\#c_2^i\.\#E^i$, we obtain the rules
\e \#c_1^r\.\#E^r = -\#c_1^i\.\#E^i, \l{ErEi1}\f
\e \#c_2^r\.\#E^r=-\#c_2^i\.\#E^i, \l{ErEi2}\f
whence there is no coupling in reflection from TE$_{c1}$ wave to TE$_{c2}$ wave, or from TE$_{c2}$ wave to TE$_{c1}$ wave. Writing the relations between the incident and reflected fields  in these two cases as
\ea \#a_1\.(\#E_1^i+ \#E_1^r) + \#b_1\.\h_o(\#H_1^i+ \#H_1^r)&=&0, \\ 
\#a_2\.(\#E_2^i+ \#E_2^r) + \#b_2\.\h_o(\#H_2^i+ \#H_2^r)&=&0, \fa
shows us that the TE$_{c1}$ wave and TE$_{c2}$ wave components actually satisfy the respective boundary conditions \r{aEbH1} and \r{aEbH2} individually. This generalizes the same property previously found for SHDB boundaries in \cite{SHDB} and for GSHDB boundaries in \cite{GSHDB}. 

In conclusion, reflection of any given incident plane wave from a boundary defined by the general conditions \r{aEbH1} and \r{aEbH2} can be solved by decomposing the incident wave in its TE$_{c1}$ wave and TE$_{c2}$ wave components and applying \r{ErEi1}, \r{ErEi2} to construct the reflected wave.

\section{Plane Waves Matched to Boundary}

By definition, a single plane wave is matched to a boundary whenever it satisfies the boundary conditions by itself. From \r{EiEr} we see that $\#E^i\not=0, \#E^r=0$ implies $J^i=0$, in which case the "incident wave" is matched to the boundary. Actually, this corresponds to an eigenvalue zero of $\=R$, whence there is no dyadic $\=R{}^{-1}$. Similarly, from \r{ErEi}, $\#E^r\not=0, \#E^i=0$ implies $J^r=0$, whence the "reflected wave" is matched to the boundary. This, again, corresponds to zero eigenvalue of $\=R{}^{-1}$, in which case there is no dyadic $\=R$. Actually, the labels "incident" and "reflected" have no special significance here, since the wave vectors of the matched waves differ only by the sign of their normal component. Surface waves and  leaky waves are two well-known examples of matched waves, which occur when $k_n$ contains an imaginary component \cite{Ishimaru}.

\subsection{Matched-Wave Conditions}

The condition for a matched wave corresponding to $J^i=\#k^i\.(\#c_1^i\x\#c_2^i) = 0$ can be expressed from \r{c1} - \r{JiJr}  as
\e \#k^i\.(\#k^i\x\#b_1-k_o\#a_1)\x(\#k^i\x\#b_2-k_o\#a_2)=0, \l{match1}\f
which, in spite of its cubic appearance, is actually quadratic in $\#k^i$, 
\ea &&  k_o\#k^i\.(\#a_1\x\#a_2+ \#b_1\x\#b_2)  - (\#k^i\x\#a_1)\.(\#k^i\x\#b_2) \nonumber\\
&+& (\#k^i\x\#b_1)\.(\#k^i\x\#a_2) =0. \l{disp}\fa
The equation for the matched reflected wave equals \r{disp} with $\#k^i$ replaced by $\#k^r$.

Substituting $\#k^i=-k_n\#n + k_t\#u_t$ with $\#u_t\.\#u_t=1$ and $k_n=\sqrt{k_o^2-k_t^2}$, \r{disp} defines a certain planar dispersion diagram $k_t = k_t(\#u_t)$ for waves matched to a given boundary. The same diagram is obtained when starting from the condition $J^r=0$. It can be conceived as a 2D counterpart to the 3D dispersion surface associated to plane waves in a homogeneous medium \cite{GSHDB}. For complex wave vectors the dispersion diagram is complex. The case of $\#k_t\.\#k_t=0$, $k_t\not=0$ must be treated separately since there is no unit vector $\#u_t$. 

Assuming $J^r\not=0$, the polarization of the matched wave defined by $\#E^r=0$ is obtained from \r{R} as 
\e \#E^i \sim \#c_1^i\x\#c_2^i = (\#k^i\x\#b_1- k_o\#a_1)\x(\#k^i\x\#b_2- k_o\#a_2). \l{Eimatch}\f
The equation \r{disp} for $\#k^i$ actually follows from $\#k^i\.\#E^i=0$.

Let us consider some special cases.

\subsection{Examples of Matched Waves}

\subsubsection*{DB Boundary and its Generalization}

The DB boundary \r{DB} in its generalized form \cite{GDB2009} is defined by choosing $\#a_{jt}=\#b_{jt}=0$ in \r{ximp1} and \r{ximp2}, in which case the boundary conditions become
\ea \#n\.(\A_1\#B+\B_1\h_o\#D) &=& 0 \l{GDB1}\\
\#n\.(\A_2\#B+ \B_2\h_o\#D)&=&0. \l{GDB2}\fa 
This corresponds to choosing $\#a_j=\#n\A_j$ and $\#b_j=\#n\B_j$ in \r{aEbH1} and \r{aEbH2}. The dispersion equation \r{disp} takes the form
\e (\A_2\B_1-\A_1\B_2)\#k_t\.\#k_t=0.  \f
For $\A_2\B_1-\A_1\B_2=0$ the two boundary conditions \r{GDB1}, \r{GDB2} become the same, which case is omitted here. Otherwise, the dispersion equation becomes $\#k_t\.\#k_t=0$. The obvious real solution is $\#k^i=-\#n k_o$. Thus, such a dispersion diagram consists of just a point at the origin. In fact, it is previously known that a wave of any polarization with normal incidence is matched to a DB boundary \cite{GSHDB}.  

More generally, a wave defined by $\#k^i = \pm k_o\#n + \#k_t' + j\#n\x\#k_t'$ is matched to the DB boundary for any tangential vector $\#k_t'$.

\subsubsection*{Impedance Boundary}

For the choice $\#a_j=\#a_{jt}$ and $\#b_j=\#b_{jt}$ the conditions \r{aEbH1} and \r{aEbH2}  are reduced to those of the impedance boundary, \r{imp1}, \r{imp2}. The condition for a matched wave \r{disp} becomes
$$ k_ok_n\#n\.(\#a_{1t}\x\#a_{2t}+\#b_{1t}\x\#b_{2t})  + k_o^ 2(\#a_{1t}\.\#b_{2t}-\#a_{2t}\.\#b_{1t}) $$
\e - k_t^2\#u_t\#u_t: (\#a_{1t}\#b_{2t}-\#a_{2t}\#b_{1t}) =0.\l{disp2}\f
Substituting $k_t^2=k_o^2-k_n^2$ and defining $\#v_t=\#n\x\#u_t$ leaves us with a quadratic equation for $k_n$,
$$ (k_n^2\#u_t\#u_t + k_o^2\#v_t\#v_t):(\#a_{1t}\#b_{2t}-\#a_{2t}\#b_{1t}) $$
\e + k_ok_n \#n\.(\#a_{1t}\x\#a_{2t} + \#b_{1t}\x\#b_{2t})=0, \l{match2}\f
which can be solved in analytic form as
\ea k_n &=& k_o(-A \pm \sqrt{A^2 -B}),  \l{eqn}\\
 A &=& \frac{\#n\.(\#a_{1t}\x\#a_{2t} + \#b_{1t}\x\#b_{2t})}{2\#u_t\#u_t:(\#a_{1t}\#b_{2t}-\#a_{2t}\#b_{1t})} \\
B &=& \frac{\#v_t\#v_t:(\#a_{1t}\#b_{2t}-\#a_{2t}\#b_{1t})}{\#u_t\#u_t:(\#a_{1t}\#b_{2t}-\#a_{2t}\#b_{1t})}.   \fa
Once $k_n(\#u_t)$ is solved from \r{eqn} for a given unit vector $\#u_t$, the tangential wave vector component is obtained as
\e  \#k_t = \#u_t \sqrt{k_o^2 - k_n^2(\#u_t)}. \f
Let us consider two particular examples of the impedance boundary.

\subsubsection*{Isotropic Impedance Boundary}

The simple isotropic impedance boundary is defined by the condition \cite{Methods}
\e \#E_t = Z_s\#n\x\#H_t, \f
where $Z_s$ is the surface impedance. Written in the form
\ea \#u_x\.\#E + Z_s\#u_y\.\#H &=&0 \\
\#u_y\.\#E - Z_s\#u_x\.\#H &=& 0, \fa
we can identify $\#a_{1t}=\#u_x$, $\#a_{2t}=\#u_y$, $\#b_{1t}=Z_s\#u_y/\h_o$ and $\#b_{2t}=-Z_s\#u_x/\h_o$ in the representations \r{aEbH1}, \r{aEbH2}. Substituting these and $\#a_{1t}\#b_{2t}-\#a_{2t}\#b_{1t}= -(Z_s/\h_o)\=I_t$ in \r{eqn} yields
\e \frac{k_n}{k_o} = \frac{\h_o^2+Z_s^2}{2\h_o Z_s}\pm \frac{\h_o^2-Z_s^2}{2\h_o Z_s}. \f
The two solutions for a matched wave are, thus,
\e k_n= k_o\h_o/Z_s,\ \ \ \ \ k_t= \frac{k_o}{Z_s}\sqrt{Z_s^2-\h_o^2}, \l{knTE}\f
\e k_n=k_oZ_s/\h_o,\ \ \ \ \ k_t= \frac{k_o}{\h_o}\sqrt{\h_o^2-Z_s^2}. \l{knTM}\f
One can show that  $\#k_t\.\#E_t^i=0$ for the first solution. Since this implies $\#n\.\#E^i=0$, the solution corresponds to a TE$_n$-polarized wave. Similarly, the second solution can be shown to correspond to a TM$_n$-polarized wave. Since a surface wave requires $\Im\{k_n\}>0$, for the TE$_n$ or TM$_n$ cases, from \r{knTE} and \r{knTM} it is seen that we must respectively have capacitive or inductive $Z_s$, a well-known fact \cite{Collin}.

\subsubsection*{Soft-and-Hard Boundary}

As an example of an anisotropic impedance boundary, let us consider the SH boundary, defined by $\#a_{1t}=\#b_{2t}=\#u_x$ and $\#a_{2t}=\#b_{1t}=0$, whence $\#a_{1t}\#b_{2t}-\#a_{2t}\#b_{1t}=\#u_x\#u_x$. Writing $\#u_t\.\#u_x=\cos\VF$ and $\#v_t\.\#u_x=\sin\VF$, \r{eqn} is reduced to
\e k_n = \pm k_o\sqrt{-\sin^2\VF/\cos^2\VF} = \pm j k_o\tan\VF,\l{SHkn} \f
and
\e k_t(\VF) = \pm k_o/\cos\VF. \f
This result implies that the dispersion diagram for the SH boundary consists of two straight lines parallel to $\#u_y$ \cite{GSHDB}. From \r{Eimatch} we obtain for the polarization $\#E^i\sim\#u_y$.

\section{Unconventional Boundary Conditions}

Let us consider examples of boundaries obeying less common form of the conditions \r{aEbH1} and \r{aEbH2}. The case $\#b_1=\#b_2=0$, corresponds to
\e \#a_1\.\#E=0,\ \ \ \ \#a_2\.\#E=0. \l{eEE}\f
Here we assume $\#a_1\x\#a_2=\#a_{12}\not=0$, whence the scalar conditions \r{eEE} can be replaced by the vector condition
\e \#a_{12}\x\#E=0. \l{a12E}\f
Such a boundary may be called the E boundary. Similarly we can define the H boundary by requiring $\#a_1=\#a_2=0$, corresponding to
\e \#b_1\.\#H=0,\ \ \ \ \#b_2\.\#H=0. \l{hHH}\f
For $\#b_1\x\#b_2=\#b_{12}\not=0$ \r{hHH} can be replaced by
\e \#b_{12}\x\#H=0. \l{b12H} \f
As a third example we may consider boundary conditions defined by $\#a_2=\#b_1=0$ in \r{aEbH1} and \r{aEbH2}, which corresponds to 
\e \#a_1\.\#E=0,\ \ \ \ \#b_2\.\#H=0, \l{aEbH}\f
and can be called the EH-boundary conditions. Let us consider some properties associated with these boundaries.

\subsection{E Boundary}

From \r{c1} -- \r{JiJr} we can write for the E boundary
\e \#c_1^r=\#c_1^i = -k_o\#a_1, \ \ \ \#c_2^r=\#c_2^i = -k_o\#a_2, \f
\e \ J^i= k_o^2(\#k^i\.\#a_{12}),\ \ \ \ \ J^r= k_o^2(\#k^r\.\#a_{12}),\f
whence the reflection dyadic \r{R} becomes similar to that of the PEC boundary, \r{RPEC},
\e \=R= \frac{1}{\#k^r\.\#a_{12}}\#k^r\x(\#a_2\#a_1-\#a_1\#a_2) = -\=I + \frac{\#a_{12}\#k^r}{\#a_{12}\.\#k^r}. \l{RE} \f
Actually, in the special case $\#n\.\#a_1=\#n\.\#a_2=0$ we have $\#a_{12}=\#n(\#n\.\#a_{12})$, and the E boundary equals the PEC boundary. 

\begin{itemize}

\item From \r{RE} one has $\=R\.\#a_{12}=0$, which means that for the polarization of the incident wave, $\#E^i\sim\#a_{12}$, there is no reflected wave. This can also be seen from the boundary conditions \r{eEE} which are satisfied for $\#E=\#E^i\sim \#a_{12}$. Thus, such an incident wave is matched to the E boundary. The matched-wave condition \r{disp} for $\#k^i$ becomes $\#k^i\.\#a_{12}=0$, which also follows from $\#k^i\.\#E^i=0$. The possible $\#k^i$ vector solutions for the dispersion equation make a circular disk of radius $k_o$ on the plane orthogonal to $\#a_{12}$, and the dispersion curve on the boundary plane equals its projection ellipse. However, this simple interpretation presumes that the vector $\#a_{12}$ is real. For the PEC boundary $\#a_{12}$ is parallel to $\#n$, whence the matched waves are lateral waves satisfying $\#k^i\.\#n=0$.

\item Because any incident wave satisfying $\#n\.\#E^i=0$ (TE$_n$ wave) satisfies $\#k^r\.\#E^i=\#k^i\.\#E^i=0$, the reflected field $\#E^r=\=R\.\#E^i = -\#E^i$ is another TE$_n$ wave. Since the total field is tangential and satisfies $\#E=\#E^i+\#E^r=0$, an E boundary can be replaced by a PEC boundary for TE$_n$ waves. Actually, one can show that \r{RE} satisfies $\=R\.(\=R+\=I)=0$, whence $\=R$ has two eigenvalues, $-1$ and $0$. The former corresponds to the PEC boundary case and, the latter, to the matched wave case. Any incident field can be decomposed in the two eigenfields as
\e \#E^i = \frac{1}{\#n\.\#a_{12}}(\#a_{12}(\#n\.\#E^i) - \#n\x(\#a_{12}\x\#E^i)), \f
when $\#n\.\#a_{12}\not=0$. Any incident wave is split in two parts, one reflecting from the E boundary as from the PEC boundary and, the other one, with no reflection. This property may have potential engineering applications.

\end{itemize}

The  H boundary can be handled through a similar analysis.

\subsection{EH Boundary}

The conditions \r{aEbH} of the EH boundary make another obvious special case of \r{aEbH1} and \r{aEbH2}, as defined by the choice $\#a_2=\#b_1=0$. Expressing \r{ximp1} and \r{ximp2} in the form
\ea \B_1\#n\.\#D + \#a_{1t}\.\E_o\#E&=&0, \\
\A_2 \#n\.\#B + \#b_{2t}\.\M_o\#H &=& 0, \fa
the EH-boundary conditions are valid to media not necessarily isotropic. Inserting \r{aEbH} in \r{c1} and \r{c2} yields
\e \#c_1^r=\#c_1^i = -k_o\#a_1,\ \ \ \#c_2^r=\#k^r\x\#b_2,\ \ \ \#c_2^i=\#k^i\x\#b_2, \f
whence
\e J^{i,r}=-k_o(\#k^{i,r}\x\#a_1)\.(\#k^{i,r}\x\#b_2). \f
Assuming $J^r \not=0$, the reflection dyadic \r{R} becomes
\e \=R = \frac{-k_o}{J^r}\#k^r\x((\#k^r\x\#b_2)\#a_1-\#a_1(\#k^i\x\#b_2)). \l{REH}\f
Let us consider some properties of the EH boundary.

\begin{itemize}

\item Applying the expansion \r{32}, we obtain
\e \#E^i = \frac{k_o}{J^i}\#k^i\x((\#k^i\x\#b_2)(\#a_1\.\#E^i)+k_o\#a_1(\#b_2\.\h_o\#H^i)), \f
whence the incident plane wave can be decomposed in the TE$_a$ part, satisfying $\#a_1\.\#E^i=0$, and the TM$_b$ part, satisfying $\#b_2\.\#H^i =0$. The reflected field is obtained from \r{ErEi} as
\e \#E^r = \frac{-k_o}{J^r}\#k^r\x((\#k^r\x\#b_2)(\#a_1\.\#E^i)+ k_o\#a_1(\#b_2\.\h_o\#H^i)). \l{ErEH}\f
Applying the transformation $\#E\ra -\h_o\#H$, $\h_o\#H\ra\#E$ and $\#a_1\lra\#b_2$, which leaves the Maxwell equations and boundary conditions invariant, we obtain
\e \h_o\#H^r = \frac{k_o}{J^r}\#k^r\x(k_o\#b_2(\#a_1\.\#E^i)- \#k^r\x\#a_1(\#b_2\.\h_o\#H^i)). \l{HrEH}\f

\item The condition for an incident wave matched to the EH boundary becomes
\e J^i= 0\ \ \ \Ra\ \ \ (\#k^i\x\#a_1)\.(\#k^i\x\#b_2) =0. \l{matchEH}\f
From \r{REH} we can notice the property
\e \=R\.(\#a_1\x(\#k^i\x\#b_2))=0, \f
whence $\#E^i\sim \#a_1\x(\#k^i\x\#b_2)$ corresponds to the polarization of a matched wave. Also, we have $\#H^i\sim \#k^i\x\#E^i \sim \#k^i\x\#b_2$. The dispersion equation \r{matchEH} for $\#k^i$ is actually obtained from $\#k^i\.\#E^i=0$. It is easy to check that the results for soft-and-hard boundary and DB boundary of Section 3.2 are obtained as two special cases of the present case. 

\item For two  special cases of the EH boundary conditions, 
\e \#a_{1t}\.\#E =0,\ \ \ \ \ \#n\.\#H=0, \l{specEH1}\f
\e \#b_{2t}\.\#H =0,\ \ \ \ \ \#n\.\#E=0, \l{specEH2}\f
the boundary can be replaced by the respective PEC and PMC boundaries. To see this, from \r{specEH1} the plane waves satisfy
\ea \#n\.k_o\h_o(\#H^r+\#H^i)  &=& \#n\.(\#k^r\x\#E^r+\#k^i\x\#E^i) \nonumber\\
&=& (\#n\x\#k_t)\.\#E=0, \fa
Assuming $\#a_{1t}\x(\#n\x\#k_t)=\#n(\#a_{1t}\.\#k_t )\not=0$, we obtain
\ea 0 &=& (\#a_{1t}\x(\#n\x\#k_t) )\x\#E \nonumber\\
&=& (\#n\x\#k_t)(\#a_{1t}\.\#E) -\#a_{1t}(\#n\x\#k_t)\.\#E \nonumber\\
&=& (\#a_{1t}\.\#k_t)\#n\x\#E, \fa
which equals the condition of the PEC boundary. Since this is a linear condition and independent of the plane wave, it is valid for any fields of the special EH boundary case \r{specEH1}. Similarly, \r{specEH2} can be shown to equal the PMC boundary condition $\#n\x\#H=0$. 

\begin{figure}
	\centering
		\includegraphics[width=7cm]{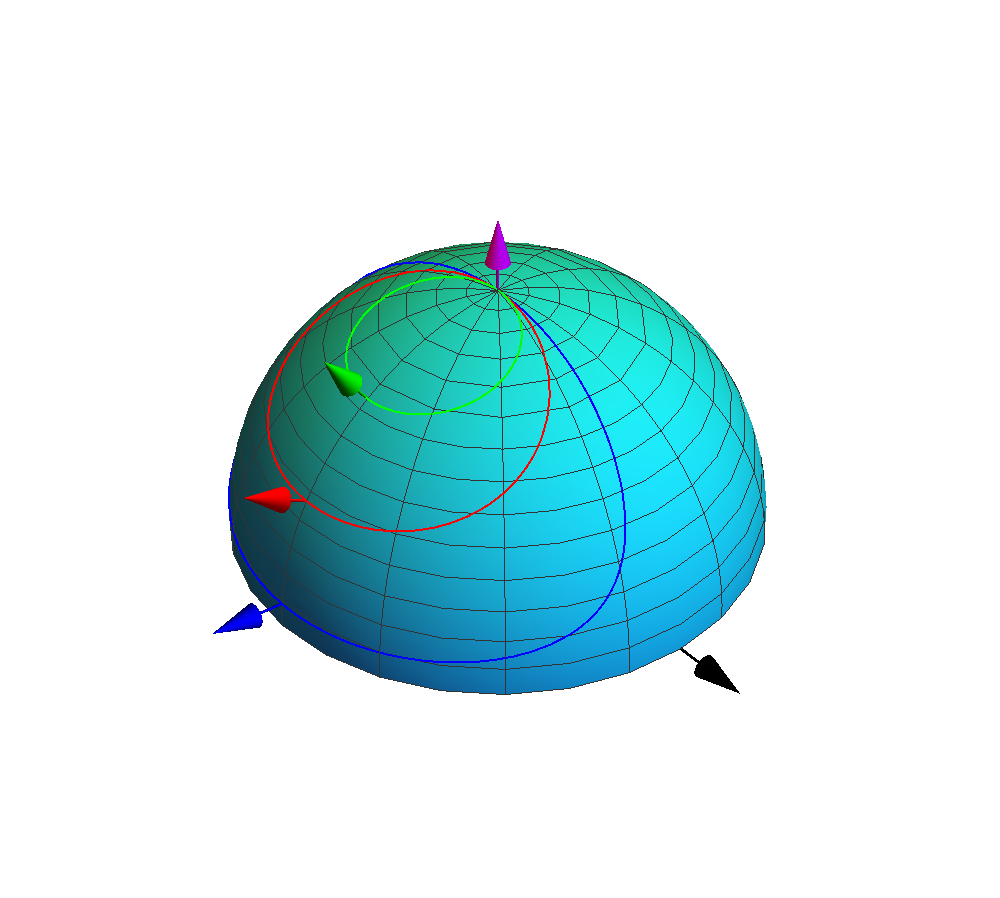}
	\label{fig:Fig1}
 \caption{Paths of the normalized matched-wave vector $\#k^i/k_o$ on the unit half sphere above the EH boundary for $\#b_2=\#n= \#u_z$ and $\#a_1$ making angles $\psi=\pi/5$, $\pi/3$ and $\pi/2-.01$ (arrows in descending order). The arrow on the right corresponds to $\#u_y\sim -\#a_1\x\#b_2$.}
\end{figure}

\begin{figure}
	\centering
		\includegraphics[width=7cm]{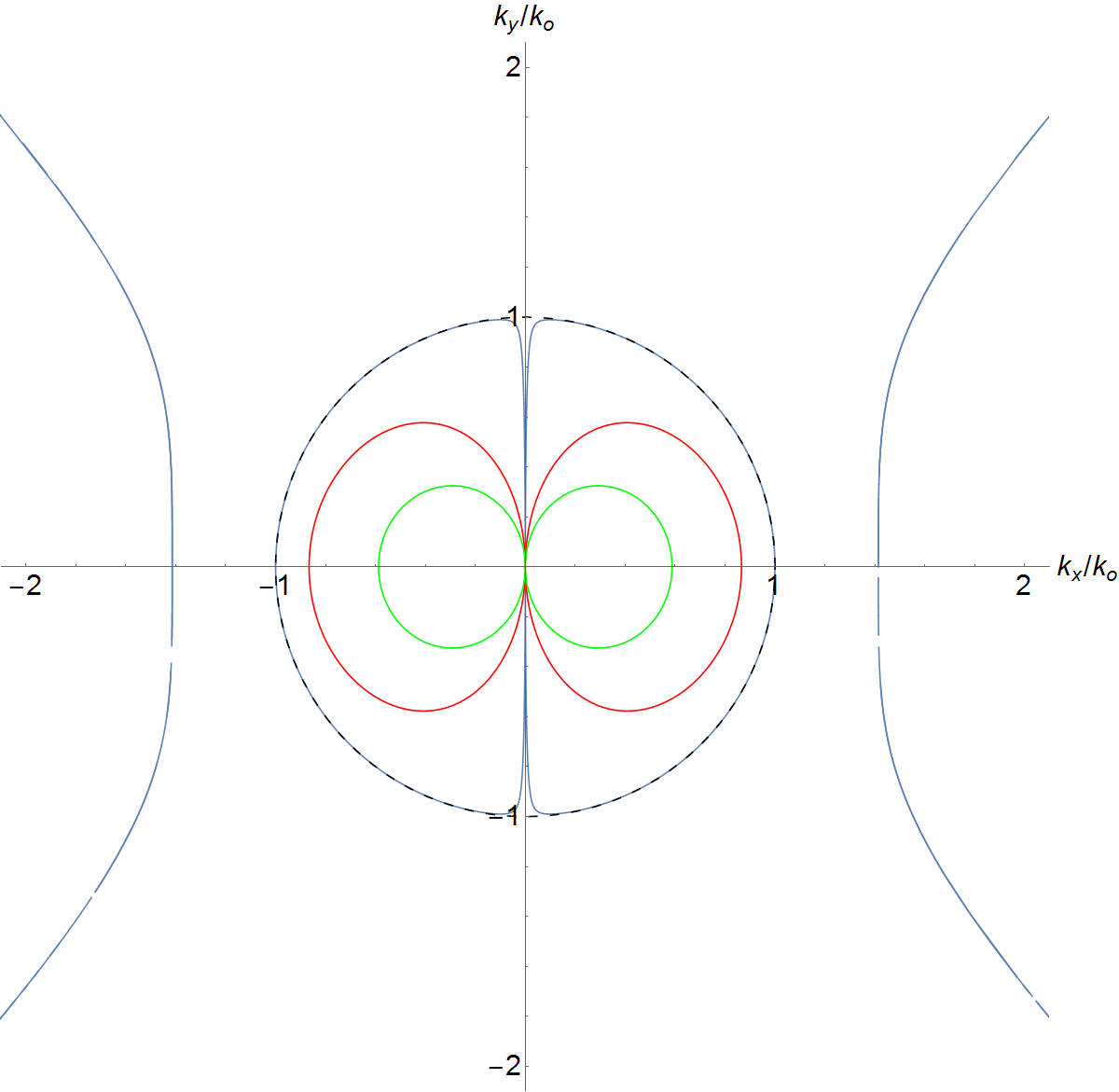}
	\label{fig:Fig2}
 \caption{Projections of the wave-vector paths on the boundary plane make the corresponding dispersion diagrams for the matched waves on different EH boundaries of Fig. 1. The lines outside the unit circle represent waves with imaginary normal component of $\#k^i$.}
\end{figure}

\item From
\e ((\#a_1\x\#b_2)\x\#a_1)\.((\#a_1\x\#b_2)\x\#b_2) = (\#a_1\x\#b_2)^2(\#a_1\.\#b_2)\f
we see that a solution $\#k^i$ of \r{matchEH} cannot be parallel to $\#a_1\x\#b_2$ unless $\#a_1$ and $\#b_2$ are orthogonal. Assuming  $(\#a_1\x\#b_2)^2\not=0$, the vectors $\#a_1, \#b_2$ and $\#a_1\x\#b_2$ form a basis and we can expand
\e \#k^i= C_1\#a_1+ C_2\#b_2 + C_{12} (\#a_1\x\#b_2). \l{kiEH} \f
Substituting this in \r{matchEH} yields a relation for the three coefficients,
\e C_1C_2= C_{12}^2 (\#a_1\.\#b_2). \f
Similarly, substituting \r{kiEH} in $\#k^i\.\#k^i=k_o^2$, another relation between $C_1, C_2$ and $C_{12}$ is obtained, whence  $C_2$ and $C_{12}$ can be expressed in terms of $C_1$. Thus, the solution $\#k^i=\#k^i(C_1)$ of \r{matchEH} depends on values of the parameter $C_1$ which move the unit vector  $\#k^i(C_1)/k_o$ along a path on the unit sphere. The dispersion diagram $\#k_t(C_1)/k_o$ is obtained as the projection of the path on the boundary plane defined by the normal vector $\#n$. 

\item As a numerical example, let us assume $\#b_2=\#n=\#u_z$ and $\#a_1 = \#u_z\cos\psi+ \#u_x\sin\psi$, whence $\#a_1\x\#b_2$ is parallel to $\#u_y$. Substituting these, and $\#k^i= k_o\#u^i$ with $\#u^i=\#u_z\cos\TH + \#u_x\sin\TH\cos\VF+ \#u_y\sin\TH\sin\VF$, in \r{matchEH}, yields the relation
\e \sin\TH(\tan\psi -\frac{\tan\TH}{\cos\VF})=0. \l{sinTH}\f
Requiring vanishing of the bracketed term and applying $k_t = \tan\TH\sqrt{k_o^2-k_t^2}$ we obtain
\e k_t=\pm\frac{k_o\sin\psi\cos\VF}{\sqrt{\cos^2\psi+ \sin^2\psi\cos^2\VF}}. \f

Figure 1 depicts the path of the direction of $\#k^i/k_o$ on the unit sphere for three different vectors $\#a_1$ at angles $\psi$. Because $-\#k^i/k_o$ is also a solution, there is another set of similar paths on the other half sphere. Fig. 2 shows the projection of both sets of paths on the plane $z=0$ of the EH boundary. There is a solution $\TH=0$ (origin in Fig.2) of \r{sinTH} for any $\psi$. For $\psi\ra\pi/2$, $\#a_1$ becomes tangential to the boundary and the dispersion diagram approaches a circle corresponding to the PEC boundary. In Fig.2 dispersion curves for the complex vector $\#a_1=j\#u_z+ \sqrt{2}\#u_x$ have been added.  Since the real part of $k_t/k_o$ stays outside the unit circle, $k_n$ is imaginary corresponding to either a surface wave or a leaky wave.

\end{itemize}

\section{Conclusion}

Analysis of plane-wave reflection from the general linear and local boundary conditions \r{ximp1} and \r{ximp2}, which were introduced in \cite{PIERL2016}, is the topic of the present paper. For the analysis the boundary surface is assumed planar and the medium above the boundary is assumed isotropic, which allows the boundary conditions to take the more compact form \r{imp1} and \r{imp2}. The field reflected from the boundary for any given incident field can be found in terms of the reflection dyadic, whose expression \r{R} is derived in Section 2. The expression is verified for known cases of PEC and soft-and-hard (SH) boundaries. It is shown that any incident wave, and the corresponding reflected wave, can be decomposed in two components, labeled as TE$_{c1}$ and TE$_{c2}$ waves, with fields orthogonal to certain vectors $\#c_1$ and $\#c_2$ depending on the boundary and the wave vector. It is also shown that the two decomposed waves do not interact in reflection, which aids in solving the reflection problem. A plane wave which satisfies the boundary conditions identically (incident wave with no reflected wave or vice versa) is called a wave matched to the boundary. Conditions for the wave vector and polarization of matched waves corresponding to given boundary parameters are found. Certain boundary conditions with known reflection properties are used as verification of the theory. The general form \r{imp1}, \r{imp2} also suggests some obvious and simple sets of boundary conditions which appear to have been omitted in the past. As three possible novelties, the E-boundary, the H-boundary and the EH-boundary, are introduced and some of their properties are studied in the present paper. In an Appendix, a suggestion of realizing the general boundary conditions by an interface of a bi-anisotropic medium is suggested. The present study concentrates on algebraic properties of fields and boundaries which is mainly of theoretical interest. However, since realizations of various special cases of boundaries defined by \r{ximp1} and \r{ximp2} as metasurfaces have been reported in \cite{AP2006} -- \cite{Zaluski14} and applications have been pointed out in \cite{Zhang} -- \cite{Frezza}, it is expected that similar work on more general boundaries will eventually be done as well. Conditions associated with reciprocity and power properties are left as  topics of forthcoming studies.

\section*{Appendix: Realization by Medium Interface}

Let us study a possible realization of the most general boundary conditions \r{ximp1}, \r{ximp2} by the interface $\#n\.\#r=0$ of a bi-anisotropic medium defined by the medium conditions \cite{Methods}
\e \am \#D\\ \#B\a = \amm \=\E & \=\xi\\ \=\z & \=\M\a \.\am \#E\\ \#H\a. \f
Assuming that the medium above the interface is isotropic with parameters $\E_o,\M_o$, from continuity of fields through the interface, the fields at the medium side of the interface must satisfy
\ea && \A_1c\#n\.(\=\z\.\#E + \=\M\.\#H) + \frac{\B_1}{\E_o}\#n\.(\=\E\.\#E+\=\xi\.\#H) \nonumber\\
&+&   \#a_{1t}\.\#E + \h_o\#b_{1t}\.\#H =0, \l{med1}\\
&&\A_2c\#n\.(\=\z\.\#E + \=\M\.\#H) + \frac{\B_2}{\E_o}\#n\.(\=\E\.\#E + \=\xi\.\#H) \nonumber\\
&+& \#a_{2t}\.\#E + \h_o\#b_{2t}\.\#H =0. \l{med2}\fa 
While it is known that there is no unique way to define a medium corresponding to given boundary conditions, let us consider a simple scenario by requiring that the conditions \r{med1} and \r{med2} are valid everywhere in the medium for any possible fields. In such a case, the medium dyadics are required to satisfy the conditions
\ea \#n\.(c\A_1\=\z+ \B_1\=\E/\E_o) + \#a_{1t} &=& 0, \\
 \#n\.(c\A_2\=\z+ \B_2\=\E/\E_o) + \#a_{2t} &=& 0, \\
  \#n\.(c\A_1\=\M+ \B_1\=\xi/\E_o) + \h_o\#b_{1t} &=& 0, \\
   \#n\.(c\A_2\=\M+ \B_2\=\xi/\E_o) + \h_o\#b_{2t} &=& 0. \fa
   
 Assuming $\De=\A_1\B_2-\A_2\B_1\not=0$,  we obtain the following restrictions to the medium dyadics,
\ea \De\#n\.\=\E &=& \E_o(\A_2\#a_{1t}-\A_1\#a_{2t}), \\
\De\#n\.\=\xi &=& \E_o\h_o(\A_2\#b_{1t}-\A_1\#b_{2t}), \\
\De\#n\.\=\z &=& \frac{1}{c}(\B_1\#a_{2t}-\B_2\#a_{1t}), \\
\De\#n\.\=\M &=& \frac{\h_o}{c}(\B_1\#b_{2t}-\B_2\#b_{1t}), \fa
whence they can be expressed as
\e \=\E =  \frac{\E_o}{\De}\#n(\A_2\#a_{1t}-\A_1\#a_{2t}) + \=\E{}',\ \ \ \ \#n\.\=\E{}'=0, \l{eps}\f
\e \=\xi =  \frac{1}{c\De}\#n(\A_2\#b_{1t}-\A_1\#b_{2t}) + \=\xi{}', \ \ \ \ \#n\.\=\xi{}'=0,\f
\e \=\z =  \frac{1}{c\De}\#n(\B_1\#a_{2t}-\B_2\#a_{1t}) + \=\z{}', \ \ \ \ \#n\.\=\z{}'=0,\f
\e  \=\M =  \frac{\M_o}{\De}\#n(\B_1\#b_{2t}-\B_2\#b_{1t}) + \=\M{}', \ \ \ \ \#n\.\=\M{}'=0.\l{mu}\f
Here, $\=\E{}' , \=\xi{}', \=\z{}', \=\M{}'$ may be any four dyadics satisfying the above conditions. One can verify that  \r{med1} and \r{med2} are satisfied when substituting \r{eps} -- \r{mu}, whence the interface acts as a boundary defined by the conditions \r{ximp1} and \r{ximp2}. The question on how the medium dyadics \r{eps} -- \r{mu} can be realized by some physical material is out of scope of this paper. 

As an example, for the GSHDB boundary with $\A_2=\B_1=0$ and $\#a_2=\#b_1=0$, the required medium dyadics have the form
\ea \=\E &=& -\E_o\frac{\#n\#a_{2t}}{\B_2}+ \=\E{}',\ \ \ \=\xi = -\sqrt{\M_o\E_o}\frac{\#n\#b_{2t}}{\B_2}+\=\xi{}',\\
 \=\z&=&-\sqrt{\M_o\E_o}\frac{\#n\#a_{1t}}{\A_1}+ \=\z{}',\ \ \ \=\M=-\M_o\frac{\#n\#b_{1t}}{\A_1}+\=\M{}'. \fa

Obviously, the realization scheme defined above breaks down for any boundary satisfying $\De=0$, for example, for the impedance boundary defined by \r{imp1} and \r{imp2}. Realization of the general impedance boundary in terms of a layer of special anisotropic medium has been described in \cite{AP2006}.


\begin{thebibliography}{99}


\bibitem{Senior} T.B.A. Senior and J.L. Volakis, {\it Approximate Boundary Conditions in Electromagnetics}, London, U.K.: IEE, 1995.

\bibitem{Hoppe} D.J. Hoppe and Y. Rahmat-Samii, {\it Impedance Boundary Conditions in Electromagnetics}, Washington, D.C.: Taylor and Francis, 1995.

\bibitem{Sievenpiper} D. Sievenpiper et al., ``High-impedance electromagnetic surfaces with a forbidden frequency band'', {\it Trans. IEEE Microwave Theory Tech.}, Vol. 47, No. 11, 2059, 1999.

\bibitem{PIERL2016} I.V. Lindell and A. Sihvola, ``Electromagnetic boundaries with PEC/PMC equivalence," {\it PIER Lett.}, Vol.61, pp.119--123. 2016.

\bibitem{DB} I.V. Lindell and A. Sihvola: ``Electromagnetic boundary condition and its realization with anisotropic metamaterial," {\it Phys.\ Rev.\ E}, Vol. 79, No.  2, 026604 (7 pages), 2009.

\bibitem{SHDB} I.V. Lindell and A. Sihvola, ``Soft-and-hard/DB boundary conditions realized by a skewon-axion medium," {\it Trans.\ IEEE Antennas Propag.}, Vol. 61, No.  2, pp. 768--774, 2013.

\bibitem{GSHDB}I.V.  Lindell  and A.~Sihvola, ``Generalized Soft-and-Hard/DB boundary," {\it IEEE Trans. Antennas Propag.}, Vol. 65, no. 1, pp. 226 -- 233, 2017.

\bibitem{SH} P.-S. Kildal, ``Artificially soft and hard surfaces in electromagnetics", {\it IEEE Trans. Antennas Propagat.}, Vol. 38, No. 10, pp. 1537--1544, Oct. 1990.

\bibitem{GSH} I.V. Lindell: ``Generalized soft-and-hard surface," {\it IEEE Trans.\ Antennas Propag}, Vol. 50, No. 7, pp. 926-929, July 2002.

\bibitem{Methods} I.V. Lindell, {\it Methods for Electromagnetic Field Analysis}, 2nd ed., New York: Wiley and IEEE Press, 1995.

\bibitem{Gibbs} J.W. Gibbs, {\it Vector Analysis}, New York: Dover, 1960.

\bibitem{MAP1996} I.V. Lindell and P.P. Puska, "Reflection dyadic for the soft and hard surface with application to the depolarising corner reflector," {\it IEE Proc. Microw.Antennas Propag.}, Vol. 143, No. 5, pp. 417 -- 421, October 1996.

\bibitem{Ishimaru} A. Ishimaru, {\it Electromagnetic Wave Propagation and Scattering}, Englewood Cliffs, N.J.: Prentice-Hall, 1991.

\bibitem{GDB2009} I.V. Lindell, H. Wall\'en and A. Sihvola, ``General electromagnetic boundary conditions involving normal field components," {\it IEEE Antennas and Wireless Propagation Letters}, Vol.8, pp. 877--880, 2009.

\bibitem{Collin} R.F. Collin, {\it Field Theory of Guided Waves}, New York: McGraw-Hill, 1960, Sec. 11.2.

\bibitem{AP2006} I.V. Lindell and A. Sihvola, ``Realization of impedance boundary," {\it Trans. IEEE Antennas Propag.}, Vol. 54, No. 12, pp. 3669--3676, 2006.

\bibitem{Caloz} Shahvarpour, A., T. Kodera, A. Parsa and C. Caloz, ``Arbitrary electromagnetic conductor boundaries using Faraday rotation in a grounded ferrite slab'' {\it IEEE Trans.\ Microwave Theory Tech.}, vol.58, no.11, pp.2781--2793, 2010.

\bibitem{Caloz13}  Caloz C. et al, ``Practical realization of perfect electromagnetic conductor (PEMC) boundaries using ferrites, magnetless non-reciprocal metamaterials (MNMs) and graphene," {\it Proc. URSI EMTS}, pp. 652--655, Hiroshima May 2013. 

\bibitem{Elmaghrabi}  El-Maghrabi, H. M., A. M. Attiya and E. A. Hashish, ``Design of a perfect electromagnetic conductor (PEMC) boundary by using periodic patches,'' {\it Prog.\ Electromag.\ Res.\ M}, vol.16, pp.159--169, 2011.

\bibitem{Zaluski11} Zaluski, D., D. Muha and S. Hrabar, ``DB boundary based on resonant metamaterial inclusions,'' {\it Metamaterials'2011}, Barcelona, October, pp.820--822, 2011.

\bibitem{Zaluski14} Zaluski, D., S. Hrabar and D. Muha, ``Practical realization of DB metasurface,'' {\it Appl.\ Phys.\ Lett.}, vol.\ 104, 234106, 2014.

\bibitem{Zhang} Zhang, B., H. Chen, B.-I. Wu and J. A. Kong, ``Extraordinary surface voltage effect in the invisibility cloak with an active device inside,'' {\it Phys.\ Rev.\ Lett.}, Vol. 100, 063904, 2008.

\bibitem{Yaghjian08} Yaghjian, A. and S. Maci ``Alternative derivation of electromagnetic cloaks and concentrators,'' {\it New J.\ Phys.}, Vol. 10, 115022, 2008; "Corrigendum", {\it ibid}, Vol. 11, 039802, 2009.

\bibitem{Yaghjian10} Yaghjian, A., ``Extreme electromagnetic boundary conditions and their manifestation at the inner surfaces of spherical and cylindrical cloaks,'' {\it Metamaterials}, Vol. 4, pp. 70--76, 2010.

\bibitem{Kildal09}  Kildal, P.-S., ``Fundamental properties of canonical soft and hard surfaces, perfect magnetic conductors and the newly introduced DB surface and their relation to different practical applications included cloaking,'' {\it Proc.\ ICEAA'09}, Torino, Italy Aug. 2009, pp. 607--610.

\bibitem{Frezza} N. Tedeschi, F. Frezza, and A. Sihvola, “On the Perfectly Matched Layer and the DB boundary condition,” {\it JOSA A}, vol. 30, pp. 1941-1946, Oct. 2013.


\end{thebibliography}
\end{document}